\documentclass[prd,aps,a4paper,superscriptaddress,twocolumn,nofootinbib]{revtex4}
\usepackage[colorlinks, linkcolor=blue, anchorcolor=blue, citecolor=blue]{hyperref}
\usepackage{hyperref}
\usepackage{graphicx}
\usepackage{color}
\usepackage{dcolumn}
\usepackage{bm}
\usepackage{slashed}
\usepackage{amsmath}
\usepackage{latexsym}
\usepackage{amssymb}
\usepackage{mathrsfs}
\usepackage{amsfonts}
\usepackage{url}
\usepackage{enumitem}
\setlist[itemize]{leftmargin=2em,itemsep=0em}

\allowdisplaybreaks
\begin{document}

\title{Refinements and corrections to the deflection angle of light in Kerr-de Sitter spacetime}

\author{Yang Huang}
\email[Email: ]{yanghuang55@outlook.com}
\affiliation{School of Physics and Electronic Science, Hunan University of Science and Technology, Xiangtan 411021, China}
\affiliation{Key Laboratory of Intelligent Sensors and Advanced Sensing Materials of Hunan Province, Hunan University of Science and Technology, Xiangtan 411021, China}

\author{Xiangyun Fu} 
\email[Corresponding author, Email: ]{xyfu@hnust.edu.cn}
\affiliation{School of Physics and Electronic Science, Hunan University of Science and Technology, Xiangtan 411021, China}
\affiliation{Key Laboratory of Intelligent Sensors and Advanced Sensing Materials of Hunan Province, Hunan University of Science and Technology, Xiangtan 411021, China}


\begin{abstract}
The deflection angle of equatorial light (the light in the equatorial plane) in Kerr-de Sitter (KdS) spacetime was previously calculated by Sultana [\href{https://journals.aps.org/prd/abstract/10.1103/PhysRevD.88.042003}{Phys. Rev. D 88, 042003 (2013)}]. However, we have identified three problems with his result: (a) Orbit problem, the orbit solution used for computing the deflection angle is inaccurate as it does not stem from the original equation of motion (EOM). (b) Position problem, assuming the source and observer are at infinity is physically unrealistic given the presence of the cosmological horizon. (c) Static problem, assuming the observer remains at rest in a static slice of spacetime ignores the expansion of the de Sitter space. In this paper, we address and correct these issues respectively by (a) deriving the orbit solution through solving the original EOM directly, (b) employing a widely accepted definition of the finite-distance deflection, and (c) adopting the Randers optical space for the spatial projection of null geodesics. Based on these corrections, we obtain a more precise and applicable result for the deflection angle of equatorial light in KdS spacetime.
\end{abstract}

\maketitle

\section{Introduction}\label{introduction}

The accelerated expansion of the Universe has been substantiated through the observation of Type Ia supernovae \cite{caldwell1998cosmological,perlmutter1999measurements,astier2005supernova}, large-scale cosmic structures \cite{abazajian2004second,abazajian2005third}, and anisotropies in the cosmic microwave background \cite{spergel2003first}. To account for this phenomenon, various dark energy models have been proposed by modifying the matter term of Einstein’s field equations \cite{sahni2000case,zlatev1999quintessence,caldwell2002phantom,feng2005dark,cai2016dark,mazumdar2001assisted,arkani2004ghost,deffayet2002accelerated,bento2004revival}, with the simplest among them being the cosmological constant ($\Lambda$) model \cite{sahni2000case}.

The gravitational deflection of light stands out as a powerful tool in astrophysics and cosmology, prompting researchers to delve into its application for exploring dark energy. In 1983, Islam asserted that the deflection of light remains unaffected by $\Lambda$ in the context of the Schwarzschild-de Sitter (SdS) spacetime, since $\Lambda$ is absent from the orbit of light \cite{islam1983cosmological}. Islam's methodology and findings are also documented in Refs.~\cite{freire2001cosmological,lake2002bending,kagramanova2006solar,finelli2007light}.

\begin{figure}[!ht]
    \centering
    \includegraphics[width=0.9\columnwidth]{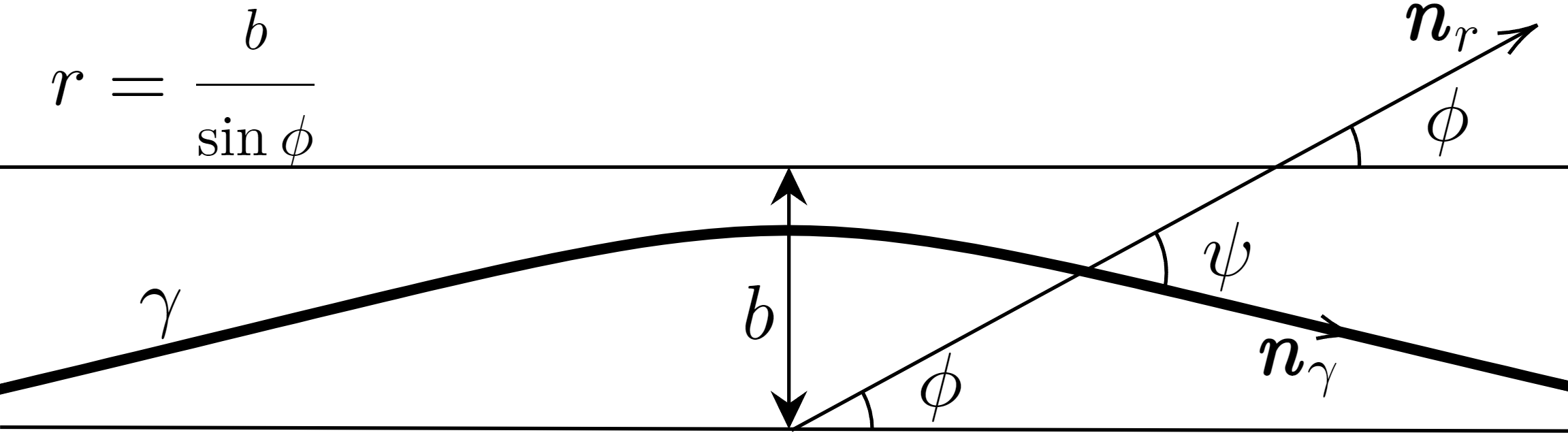}
    \caption{The schematic for the deflection angle in RI method. $\gamma$ is the trajectory of the equatorial light, the one-sided deflection angle is $\psi-\phi$. }
    \label{fig-1}
  \end{figure}

Islam's conclusion is not changed until the influence of the metric itself (which determines the actual observations) is taken into account by Rindler and Ishak (RI method) \cite{rindler2007contribution}. For the SdS spacetime whose metric states
\begin{equation}
    \mathrm{d} s^{2} =-w(r)\mathrm{d} t^{2} +\frac{1}{w(r)}\mathrm{d} r^{2} +r^{2}\mathrm{d} \theta ^{2} +r^{2}\sin^{2} \theta \mathrm{d} \phi ^{2},
    \label{SdSmetric}
\end{equation}
where $w(r)=1-2M/r-\Lambda r^2/3$, Rindler and Ishak considered the equatorial light ($\theta=\pi/2, \mathrm{d}\theta=0$) and constructed a two-dimensional space by setting $\mathrm{d}t=0$,
\begin{equation}
    \mathrm{d} l^{2} =h_{ij}\mathrm{d}x^i \mathrm{d}x^j=\frac{1}{w(r)}\mathrm{d} r^{2} +r^{2}\mathrm{d} \phi ^{2}.
    \label{RIspace}
\end{equation}
As shown in Fig.~\ref{fig-1},
in the space determined by Eq.~\eqref{RIspace}, $\gamma$ is the trajectory of light, $\boldsymbol{n}_\gamma$ is the tangent vector along $\gamma$, $\boldsymbol{n}_r$ is the outward radial vector. $\phi$ is the azimuthal coordinate, $\psi$ is the angle between $\boldsymbol{n}_r$ and $\boldsymbol{n}_\gamma$. The $\psi$ is expressed in terms of $r$ and $\phi$ through $\cos \psi =\left( \boldsymbol{n}_r \boldsymbol{\cdot} \boldsymbol{n}_\gamma\right)/\left(| \boldsymbol{n}_r| | \boldsymbol{n}_\gamma| \right) =\left( h_{ij} n_{r}^{i} n_{\gamma }^{j}\right)/\left(\sqrt{h_{ij} n_{r}^{i} n_{r}^{j}} \sqrt{h_{ij} n_{\gamma }^{i} n_{\gamma }^{j}}\right)$. By solving the EOM, Rindler and Ishak derived the orbit solution, i.e., the radial coordinate of $\gamma$ is expressed in term of the azimuthal coordinate
\begin{equation}
    u\approx\frac{\sin \phi}{b}+M\cdot \frac{\cos(2\phi)+ 3 }{2 b^2} ,
    \label{SdSsolution}
\end{equation}
where $b$ is the impact parameter, and the variable replacement $u=1/r$ is adopted throughout this paper. Then for the source and observer "far" from the lens, the deflection angle is calculated by
\begin{equation}
    \delta = \left. 2(\psi-\phi)\right|_{\phi=0}.
    \label{rideflectionangle}
\end{equation}
Specifically, when $\phi=0$, one has $r=r_0\equiv r(\phi=0)$, accordingly, $\psi=\psi(r=r_0,\phi=0)$, thus the deflection angle of equatorial light in SdS spacetime is obtained as
\begin{equation}
    \delta \approx M\cdot \frac{4}{b} -M^{3} \cdot \frac{8}{b^{3}} -\frac{\Lambda }{M} \cdot \frac{b^{3}}{6},
    \label{DASdSRI}
\end{equation}
which contains the $\Lambda$.

On one hand, certain researchers have embraced the RI method as an alternative approach to investigating the deflection of light \cite{bhattacharya2010light,bhattacharya2011vacuole,sultana2012bending,sultana2013contribution,farrugia2016solar,ali2018light,mishra2018trajectories,seccuk2020bending,he2020gravitational}. On the other hand, this method has some problems and encounters some criticisms from scholars \cite{park2008rigorous,simpson2010lensing,ishihara2016gravitational,lake2002bending,bhadra2010gravitational,arakida2012effect}. By using the RI method, Sultana calculated the deflection angle of the equatorial light in KdS spacetime whose metric reads \cite{bardeen1973houches}
\begin{equation}
    \begin{aligned}
        \mathrm{d} s^{2} = & \frac{a^{2} \Delta _{\theta }\sin^{2} \theta -\Delta _{r}}{Q^{2} \Sigma }\mathrm{d} t^{2} +\frac{\Sigma }{\Delta _{r}}\mathrm{d} r^{2} +\frac{\Sigma }{\Delta _{\theta }}\mathrm{d} \theta ^{2}\\
         & +\frac{\sin^{2} \theta }{Q^{2} \Sigma }\left[ \Delta _{\theta }\left( a^{2} +r^{2}\right)^{2} -a^{2} \Delta _{r}\sin^{2} \theta \right]\mathrm{d} \phi ^{2}\\
         & -\frac{2a\sin^{2} \theta }{Q^{2} \Sigma }\left[ \Delta _{\theta }\left( a^{2} +r^{2}\right) -\Delta _{r}\right] \mathrm{d} t\mathrm{d} \phi ,
        \end{aligned}
    \label{kdsmetric0}
    \end{equation}
where
\begin{equation}
    \begin{aligned}
        Q  =&1+\frac{\Lambda a^2}{3},\ \ \ \Delta_r  =\left(1-\frac{\Lambda r^2}{3}\right)\left(r^2+a^2\right)-2 M r, \\
        \Sigma  =&r^2+a^2 \cos ^2 \theta,\ \ \ \Delta_\theta  =1+\frac{\Lambda a^2 \cos ^2 \theta}{3}.\\
        \end{aligned}
        \label{kdsmetric1}
    \end{equation}
    The outcome of his calculations is expressed as \cite{sultana2013contribution},
\begin{equation}
\begin{aligned}
    \delta =  & M\cdot \frac{4}{b} +\Lambda \cdot \frac{5\pi b^{2}}{32} +M^{2} \cdot \frac{15\pi }{4b^{2}} -Ma\cdot \frac{4}{b^{2}}\\
     & -M\Lambda \cdot b\left(\frac{75\pi ^{2}}{512} +\frac{101}{96}\right) +a\Lambda \cdot \frac{5\pi b}{16}\\
     & -\frac{a\Lambda }{M} \cdot \frac{b^{2}}{6} -\frac{\Lambda }{M} \cdot \frac{b^{3}}{6} +\mathcal{O}\left( \Lambda ^{2}\right) ,
    \end{aligned}
    \label{DAKdSSultana}
\end{equation}
which keeps only second order terms in parameters $M/b$, $a/b$, $\Lambda b^2$, and $\phi_0$. $\phi_0\ll 1$ is the approximate azimuthal coordinate of the source or observer. Sultana's result naturally carries the intrinsic limitations of the RI method. What's more, when $a=0$, unexpectedly, Eq.~\eqref{DAKdSSultana} (the deflection angle in KdS spacetime) fails to reduce to Eq.~\eqref{DASdSRI} (the deflection angle in SdS spacetime), despite both results are computed using the RI method, and Eq.~\eqref{kdsmetric0} (the metric of KdS spacetime) can be simplified to Eq.~\eqref{SdSmetric} (the metric of SdS spacetime).

In this paper, we identify and correct the deflection angle of the equatorial light in KdS spacetime computed by Sultana from three aspects: orbit problem, position problem, and static problem. In addition, our calculation process is based on the Gauss-Bonnet theorem and is commonly referred to as the Gibbons-Werner (GW) method—a widely employed approach for computing the deflection angle of particles in recent years \cite{gibbons2008applications}.

This paper is structured as follows: In Sec.~\ref{3problems}, we elaborate the orbit problem, position problem, and static problem associated with the existing result for the deflection angle of the equatorial light in KdS spacetime obtained using the RI method \cite{sultana2013contribution}, and present corresponding coping strategies. Moving on to Sec.~\ref{CalDefAngle}, building upon the framework established in the previous section, we calculate the deflection angle of the equatorial light in KdS spacetime using the GW method. Sec.~\ref{conclusion} offers a thorough summary. Throughout this paper, we adopt the geometric unit ($G=c=1$), and the spacetime signature is designated as ($-,+,+,+$).

\section{three problems}
\label{3problems}
\subsection{Orbit problem}
The original EOM of the equatorial light in KdS spacetime is a first-order differential equation. However, in Ref.~\cite{sultana2013contribution}, the orbit solution is obtained by solving a second-order differential equation derived from the original EOM, instead of directly solving the original EOM. This concern is also raised in Refs.~\cite{lake2002bending,bhadra2010gravitational,arakida2012effect}. To solve the orbit problem, we employ an iterative perturbation method proposed by Crisnejo $et\ al.$ \cite{crisnejo2019higher} to directly solve the original EOM.

\subsubsection{Original EOM}
For the equatorial light in KdS spacetime Eq.~\eqref{kdsmetric0}, with two conserved quantities (energy $E$ and angular momentum $L$), one can obtain the original EOM \cite{sultana2013contribution}
\begin{equation}
    \begin{aligned}
        \left(\frac{\mathrm{d} u}{\mathrm{d} \phi }\right)^{2} = & \frac{1-b^{2} u^{2}}{b^{2}} +M\cdot 2u^{3} +\Lambda \cdot \frac{1}{3} -Ma\cdot \frac{4u}{b^{3}}\\
         & +a^{2} \cdot \frac{3u^{2} -2b^{2} u^{4}}{b^{2}} -a\Lambda \cdot \frac{2}{3b^{3} u^{2}} + \mathcal{O} \left( \epsilon^3\right),
        \end{aligned}
        \label{dudphikds}
\end{equation}
in which $b=L/E$ is the impact parameter, and $\epsilon$ denotes the terms arising from any combination of $M/b$, $a/b$ and $\Lambda b^2$.

\subsubsection{Sultana's orbit solution}
A second-order differential equation can be obtained by differentiating Eq.~\eqref{dudphikds} with respect to $\phi$
\begin{equation}
    \begin{aligned}
    \frac{\mathrm{d}^{2} u}{\mathrm{d} \phi ^{2}} = &-u+M\cdot 3u^{2} -Ma\cdot \frac{2}{b^{3}} +a^{2} \cdot \frac{3u-4u^{3} b^{2}}{b^{2}}  \\
    &  +\Lambda a\cdot \frac{2}{3b^{3} u^{3}} +\mathcal{O}\left(\epsilon^3\right).
    \label{eomSultana}
    \end{aligned}
\end{equation} 
By solving Eq.~\eqref{eomSultana}, Sultana got the so-called orbit solution
\begin{equation}
    \begin{aligned}
        u= & \frac{\sin \phi }{b} +M\cdot \frac{\cos (2\phi )+3}{2b^{2}}\\
         & -M^{2} \cdot \frac{3[\sin (3\phi )+20\phi \cos \phi -10\pi \cos \phi ]}{16b^{3}}\\
         & -Ma\cdot \frac{2}{b^{3}} -a^{2} \cdot \frac{\sin (3\phi )}{8b^{3}} \\
         & +a\Lambda \cdot \frac{\cos (2\phi )\csc \phi }{3} +\mathcal{O}\left(\epsilon^3\right),
        \end{aligned}
        \label{uofphiSultana}
\end{equation}
which is used to extract information about the deflection angle, as detailed in Ref.~\cite{sultana2013contribution}. Although Eq.~\eqref{uofphiSultana} satisfies the second-order differential equation~\eqref{eomSultana}, it is not the solution to the original EOM~\eqref{dudphikds}. Let's demonstrate this discrepancy. Firstly, by differentiating Eq.~\eqref{uofphiSultana} with respect to $\phi$ and squaring the result, we get
\begin{equation}
    \begin{aligned}
        \left(\frac{\mathrm{d} u}{\mathrm{d} \phi }\right)^{2} = & \frac{\cos^{2} \phi }{b^{2}} -M\cdot \frac{2\sin (2\phi )\cos \phi }{b^{3}}\\
         & -M^{2} \cdot \frac{\cos \phi }{8b^{4}} \Big[ 30(\pi -2\phi )\sin \phi +52\cos \phi \\
         & +17\cos (3\phi )\Big] -a^{2} \cdot \frac{3\cos (3\phi )\cos \phi }{4b^{4}}\\
         & +a\Lambda \cdot \frac{2[\cos (2\phi )-2]\cot^{2} \phi }{3b} +\mathcal{O}\left( \epsilon ^{3}\right).
        \end{aligned}
        \label{dudphiSultanaTest}
\end{equation}
Secondly, we derive the inverse solution of Eq.~\eqref{uofphiSultana}
\begin{equation}
    \begin{aligned}
        \phi = & \arcsin (bu)+M\cdot \frac{b^{2} u^{2} -2}{b\sqrt{1-b^{2} u^{2}}}\\
         & +M^{2} \cdot \left[\frac{bu^{3}}{2\left( 1-b^{2} u^{2}\right)^{3/2}} +\frac{3bu^{3}}{4\sqrt{1-b^{2} u^{2}}}\right. \\
         & \left. -\frac{23u}{16b\sqrt{1-b^{2} u^{2}}} -\frac{15\arccos( bu)}{4b^{2}}\right]\\
         & +Ma\cdot \frac{2}{b^{2}\sqrt{1-b^{2} u^{2}}} -a^{2} \cdot \frac{u\left( 4b^{2} u^{2} -3\right)}{8b\sqrt{1-b^{2} u^{2}}}\\
         & +a\Lambda \cdot \frac{2b^{2} u^{2} -1}{3u\sqrt{1-b^{2} u^{2}}} +\mathcal{O}\left( \epsilon ^{3}\right) ,  \quad \left(\left|\phi\right|<\frac{\pi}{2}\right).
        \end{aligned}
        \label{phiofuSultana}
\end{equation}
Finally, substituting Eq.~\eqref{phiofuSultana} into Eq.~\eqref{dudphiSultanaTest} leads to
\begin{equation}
    \begin{aligned}
        \left(\frac{\mathrm{d} u}{\mathrm{d} \phi }\right)^{2} = & \frac{1-b^{2} u^{2}}{b^{2}} +M\cdot 2u^{3} -M^{2} \cdot \frac{37}{8b^{4}}\\
         & -Ma\cdot \frac{4u}{b^{3}} +a^{2}\left(\frac{3u^{2} -2b^{2} u^{4}}{b^{2}} -\frac{3}{4b^{4}}\right)\\
         & -a\Lambda \cdot \frac{2}{3b^{3} u^{2}} + \mathcal{O}\left(\epsilon^3 \right),
        \end{aligned}
\end{equation}
in which $\left(\mathrm{d}u/\mathrm{d}\phi\right)^2$ is expressed in terms of $r$ (or $u$) with the form like Eq.~\eqref{dudphikds}. Obviously, the above formula is not consistent with the original EOM~\eqref{dudphikds}. Hence the orbit solution Eq.~\eqref{uofphiSultana}, obtained and utilized by Sultana, does not adhere to the original EOM and is inaccurate for the equatorial light in KdS spacetime. Moreover, as anticipated, the second-order derivative of Eq.~\eqref{uofphiSultana} with respect to $\phi$ aligns entirely with Eq.~\eqref{eomSultana}.

\subsubsection{The revised orbit solution}
By directly solving the original EOM~\eqref{dudphikds}, we obtain the orbit solution with the method adopted in Sec.~VI.A of Ref.~\cite{crisnejo2019higher}. We assume the form of $u(\phi)$ as
\begin{equation}
    \begin{aligned}
        u( \phi ) = & u_{0}( \phi ) +M\cdot u_{1}( \phi ) +a\cdot u_{2}( \phi ) +\Lambda \cdot u_{3}( \phi )\\
         & +M^{2} \cdot u_{4}( \phi ) +Ma\cdot u_{5}( \phi ) +M\Lambda \cdot u_{6}( \phi )\\
         & +a^{2} \cdot u_{7}( \phi ) +a\Lambda \cdot u_{8}( \phi ) +\Lambda ^{2} \cdot u_{9}( \phi ) + \mathcal{O}\left(\epsilon^3\right).
        \end{aligned}
        \label{uofphi01}
\end{equation}
Then squaring the first-order derivative of Eq.~\eqref{uofphi01} with respect to $\phi$, and comparing the result with Eq.~\eqref{dudphikds} term by term, we obtain
\begin{equation}
    \begin{aligned}
        u_{0} = & \frac{\sin \phi }{b} , & u_{1} = & \frac{\cos^{2} \phi +1}{b^{2}} ,\qquad  u_{2} =0,\\
        u_{3} = & \frac{b\sin \phi }{6} , & u_{4} = & \frac{20\sin \phi -[30\phi +3\sin (2\phi )]\cos \phi }{8b^{3}} ,\\
        u_{5} = & -\frac{2}{b^{3}} , & u_{6} = & \frac{[8\cos \phi +8(\sec \phi -1)]\cos \phi }{24} ,\\
        u_{7} = & \frac{\sin^{3} \phi }{2b^{3}} , & u_{8} = & \frac{\cos (2\phi )\csc \phi }{3} ,\quad  u_{9} =-\frac{b^{3}\sin \phi }{72} .
        \end{aligned}
        \label{unofphi}
\end{equation}
Consequently, the orbit solution for the equatorial light in KdS spacetime is presented with Eqs.~\eqref{uofphi01} and \eqref{unofphi}. In the subsequent portions of this paper, any reference to Eq.~\eqref{uofphi01} denotes the amalgamation of both Eqs.~\eqref{uofphi01} and \eqref{unofphi}.

To verify our orbit solution and support the related calculations in Sec.~\ref{CalDefAngle}, we additionally derive the inverse solution of Eq.~\eqref{uofphi01}
\begin{equation}
    \phi(u) = \begin{cases}
    \Phi(u), & \text{if  } \left| \phi \right| <\frac{\pi}{2}, \\
    \pi - \Phi(u) , & \text{if  } \left| \phi \right| >\frac{\pi}{2},
        \end{cases}
        \label{phigamma}
\end{equation}
where
\begin{equation}
    \begin{aligned}
        \Phi (u)= & \arcsin (bu)+M\cdot \frac{b^{2} u^{2} -2}{b\sqrt{1-b^{2} u^{2}}} -\Lambda \cdot \frac{b^{3} u}{6\sqrt{1-b^{2} u^{2}}}\\
         & +M^{2} \cdot \left[\frac{u\left( 20b^{2} u^{2} -3b^{4} u^{4} -15\right)}{4b\left( 1-b^{2} u^{2}\right)^{3/2}}\right. \\
         & \left. +\frac{15\arcsin (bu)}{4b^{2}}\right] +Ma\cdot \frac{2}{b^{2}\sqrt{1-b^{2} u^{2}}}\\
         & +M\Lambda \cdot \frac{b}{6\left( b^{2} u^{2} -1\right)^{2}}\bigg[ 2b^{4} u^{4} -2\sqrt{1-b^{2} u^{2}} +2 \\
         &  +b^{2} u^{2}\left( 3\sqrt{1-b^{2} u^{2}} -4\right)\bigg] -a^{2} \cdot \frac{bu^{3}}{2\sqrt{1-b^{2} u^{2}}}\\
         & +a\Lambda \cdot \frac{\left( 2b^{2} u^{2} -1\right)}{3u\sqrt{1-b^{2} u^{2}}} +\Lambda ^{2} \cdot \frac{b^{5} u\left( 3-2b^{2} u^{2}\right)}{72\left( 1-b^{2} u^{2}\right)^{3/2}} \\
         &+\mathcal{O}\left( \epsilon ^{3}\right).
        \end{aligned}
        \label{PHIofu}
\end{equation}
For the first-order derivative of Eq.~\eqref{uofphi01} with respect to  $\phi$, substituting Eq.~\eqref{phigamma} into it and then squaring the result yields a formula that is entirely congruent with the original EOM~\eqref{dudphikds}, which validates the correctness of our orbit solution.

\subsection{Position problem}
As depicted in Eq.~\eqref{rideflectionangle}, the original RI method calculates the deflection angle by setting the azimuthal coordinate to zero, implying the source and observer are located at infinity. Similarly, in Ref.~\cite{sultana2013contribution}, the positions of the source and observer are determined under the approximation that the azimuthal coordinate is "sufficiently small". However, in both SdS and KdS spacetimes, the source and observer cannot reach infinity due to the existence of the cosmological horizon (or de Sitter horizon). Criticisms of the position problem in the RI method are also articulated in Refs.~\cite{bhadra2010gravitational,ishihara2016gravitational}. To solve the position problem, we adopt a definition of finite-distance deflection angle proposed by Ishihara $et\ al.$ \cite{ishihara2016gravitational}.

\subsubsection{The position problem in SdS and KdS spacetimes}
Considering the SdS spacetime Eq.~\eqref{SdSmetric}, for $0<9 \Lambda M^{2}<1$, there exist two positive roots $r_{+}$ and $r_{++}$ of $w(r)$ such that $0<2M<r_{+}<3M<r_{++}$. The root $r_{+}=(2 / \sqrt{\Lambda}) \cos (\epsilon / 3+4 \pi / 3)$, with $\cos \epsilon=-3 M\sqrt{\Lambda}$, describes the event horizon, and the root $r_{++}=(2 / \sqrt{\Lambda}) \cos (\epsilon / 3)\approx \sqrt{3/\Lambda}$ localizes the cosmological horizon \cite{podolsky1999structure}. As a consequence, the source and observer cannot exceed the cosmological horizon. When $\phi=0$, the assumption of the infinite-distance source and observer is physically unrealistic, although the orbit solution Eq.~\eqref{SdSsolution} is mathematically reasonable.

Considering the KdS spacetime Eq.~\eqref{kdsmetric0}, it also has a cosmological horizon at $r\approx \sqrt{3/\Lambda}$. When $\phi=0$, not only is the assumption of the infinite-distance source and observer physically unrealistic, but also the orbit solution given by Eq.~\eqref{uofphiSultana} becomes mathematically singular (divergent). Therefore, in \cite{sultana2013contribution}, Sultana slightly modified the procedure of the RI method—the positions of the source and observer are determined by employing the approximation $\phi \ll 1$. This approximation is unclear and cannot guarantee that the source and observer are located within the cosmological horizon.

As mentioned previously, despite the same method is utilized, the result for the KdS case (Eq.~\eqref{DASdSRI}) does not reduce to that for the SdS case (Eq.~\eqref{DAKdSSultana}) when $a=0$. We speculate that it is the different schemes for determining the source and observer's location, i.e., $\phi=0$ for SdS spacetime and $\phi\ll 1$ for KdS spacetime, that results in the incompatibility between the results of two cases. Additionally, the position problem is indeed noticed in Refs.~\cite{rindler2007contribution,sultana2013contribution}, but the accompanying explanations are somewhat ambiguous.

\subsubsection{Finite-distance deflection angle}
Now that the assumption of the infinite-distance source and observer is inappropriate for calculating the deflection angle of light in KdS spacetime, a logical alternative is to contemplate a finite-distance deflection angle. In 2017, Ishihara $et\ al.$ introduced a finite-distance deflection angle for the light in curved spacetimes \cite{ishihara2016gravitational} and demonstrated it is geometric invariant, i.e. well-defined, by using the Gauss-Bonnet theorem. 

\begin{figure}[!ht]
    \centering
    \includegraphics[width=0.9\columnwidth]{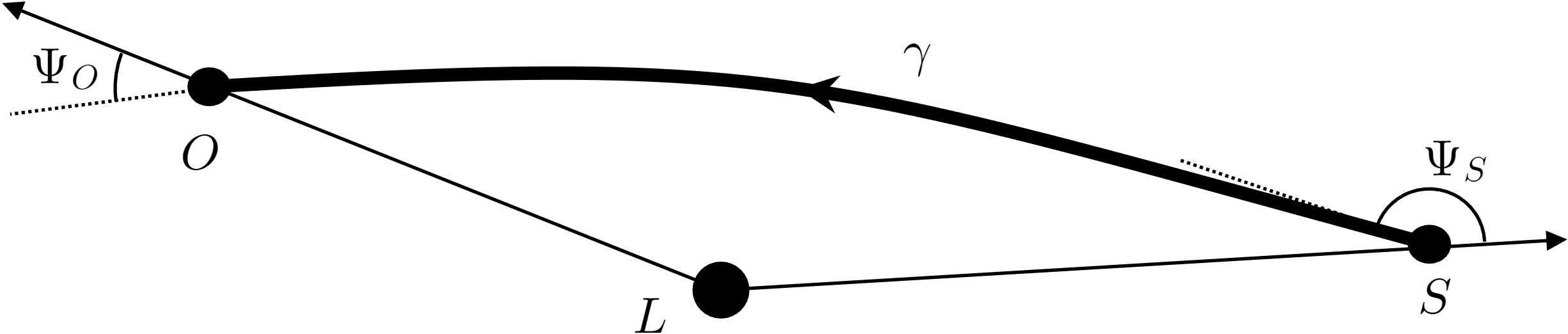}
    \caption{The schematic for the finite-distance deflection angle. $L$ is the lens, $\gamma$ is the trajectory of the light from the source $S$ to the observer $O$. $\Psi_S$ and $\Psi_O$ represent the angles between the outward radial direction and the tangent along $\gamma$ at $S$ and $O$, respectively.}
    \label{fig-2}
  \end{figure}

As shown in Fig.~\ref{fig-2}, 
$L$ is the lens, $\gamma$ is the trajectory of the equatorial light from the source $S$ to the observer $O$. $\Psi_S$ and $\Psi_O$ represent the angles between the outward radial direction and the tangent along $\gamma$ at $S$ and $O$, respectively. Then the finite-distance deflection angle is defined as \cite{ishihara2016gravitational}
\begin{equation}
    \delta = \Psi_O- \Psi_S + \phi_{OS},
    \label{defineangle}
\end{equation}
where $\phi_{OS}=\phi_O-\phi_S$ is the increment of the azimuthal coordinate. $\phi_S$ and $\phi_O$ are the azimuthal coordinate of $S$ and $O$, respectively, and $\phi_S<\pi/2<\phi_O$ is assumed without losing generality. When the $S$ and $O$ approach the infinity, $\Psi_S=\pi$ and $\psi_O=0$, accordingly Eq.~\eqref{defineangle} reduce to the conventional infinite-distance deflection angle. The definition given by Eq.~\eqref{defineangle} has been extensively utilized in studying the deflection of massless and massive particles \cite{ishihara2017finite,ono2017gravitomagnetic,ono2018deflection,ono2019deflection,haroon2019shadow,kumar2019shadow,crisnejo2019finite,ono2019effects,li2020finite,li2020thefinitedistance,takizawa2020gravitational,li2020circular,li2021kerr,li2021deflection,huang2022generalized,belhaj2022light,li2021kerrnewman,li2022deflection,pantig2022testing,huang2023finite,huang2023extending}. In this paper, it will be used to calculate the deflection angle of equatorial light in KdS spacetime to address the position problem.

\subsection{Static problem}
In Ref.~\cite{sultana2013contribution}, like the original RI method, the reduced space constructed for discussing the light deflection is derived with $\mathrm{d}t=0$, indicating that the observer is assumed to be at rest in a static slice of spacetime. However, in reality, the observer co-moves with the expanding de Sitter space. Further discussion about this issue can be found in Refs.~\cite{park2008rigorous,simpson2010lensing,ishihara2016gravitational}. To solve the static problem, we discuss the light deflection in the Randers optical space, since the geodesic of such space can be used to describe the spatial projection of null geodesics in stationary axially symmetric (SAS) spacetimes, which is guaranteed by Fermat's principle.

\subsubsection{Randers optical space}
The metric of the SAS spacetimes can be expressed as
\begin{equation}
    \begin{aligned}
    \mathrm{d} s^{2} = & g_{\mu \nu } (r,\theta )\mathrm{d} x^{\mu }\mathrm{d} x^{\nu } \\
    = &g_{tt}\mathrm{d} t^{2} +2g_{t\phi }\mathrm{d} t\mathrm{d} \phi +g_{rr}\mathrm{d} r^{2} +g_{\theta \theta }\mathrm{d} \theta ^{2} +g_{\phi \phi }\mathrm{d} \phi ^{2}.
    \end{aligned}
    \label{SASMetric}
\end{equation}
The Randers optical space ($M^{(RO)}$) for an SAS spacetime is defined by the following metric \footnote{Different from the Riemannian metric, which is a quadratic form, Eq.~\eqref{opticalRandersSAS} belongs to the category of Randers metric—a specific type of Finsler metric that includes an extra one-form \cite{randers1941asymmetrical}.} \cite{werner2012gravitational,ono2017gravitomagnetic} 
\begin{equation}
    \mathrm{d} \tilde{l}  = \sqrt{\alpha_{ij}\mathrm{d}x^i\mathrm{d}x^j} +\beta_i \mathrm{d}x^i,
    \label{opticalRandersSAS}
  \end{equation}
  where
\begin{align}
     \alpha_{ij}\mathrm{d}x^i \mathrm{d}x^j =& \alpha_{rr}\mathrm{d} r^2 + \alpha_{\theta\theta} \mathrm{d}\theta^2 + \alpha_{\phi\phi}\mathrm{d}\phi^2 \label{alphaijSAS} \\
     =&  \frac{g_{rr}}{-g_{tt}}\mathrm{d} r^{2} +\frac{g_{\theta\theta}}{-g_{tt}}\mathrm{d} \theta ^{2} +\frac{g_{t\phi}^{2}-g_{tt} g_{\phi\phi}}{g_{tt}^{2}}\mathrm{d} \phi ^{2} , \nonumber \\
     \beta_i \mathrm{d}x^i = & \frac{g_{t\phi}}{-g_{tt}} \mathrm{d} \phi. \label{betaSAS}
\end{align}
The spatial projection of a null geodesic in the four-dimensional spacetime Eq.~\eqref{SASMetric} encodes the information about the light orbit, including the co-movement of the observer; and such projection can be described as the geodesic in $M^{(RO)}$ \cite{werner2012gravitational}. Thus the light deflection can be studied with the help of $M^{(RO)}$. 

Eq.~\eqref{alphaijSAS} is the Riemannian part of the Randers optical metric Eq.~\eqref{opticalRandersSAS}, and we designate the three-dimensional space determined by it as $M^{(\alpha 3)}$. One can deduce that a geodesic in $M^{(OP)}$ can be put in one-to-one correspondence with a curve, denoted by $\gamma$, in $M^{(\alpha 3)}$ \cite{ono2017gravitomagnetic}. The deviation between $\gamma$ and the geodesic in $M^{(\alpha 3)}$ can be described by the one-form $\beta_i$. Therefore, the light deflection can also be studied within $M^{(\alpha 3)}$. 

\subsubsection{Randers optical metric of KdS spacetime}
Given the complexity of the Randers optical metric, our practical calculations are carried out within a Riemannian space, i.e. $M^{(\alpha 3)}$. Specifically, for the light confined to the equatorial plane of KdS spacetime, the corresponding $M^{(\alpha 3)}$ reduces to a two dimensional space (denoted by $M^{(\alpha 2)}_{kds}$). Substituting $\theta=\pi/2$, $\mathrm{d}\theta=0$ and Eq.\eqref{kdsmetric0} into Eq.~\eqref{opticalRandersSAS} yields the metric of $M^{(\alpha 2)}_{kds}$
\begin{equation}
    \mathrm{d}l^2 = \alpha_{rr}\mathrm{d}r^2 + \alpha_{\phi\phi}\mathrm{d}\phi^2,
    \label{malpha2kds}
\end{equation}
in which
\begin{equation}
    \begin{aligned}
        \alpha _{rr} = & \frac{r^{3}\left( a^{2} \Lambda +3\right)^{2}}{\mathcal{Z}\left( r\mathcal{Z} -3a^{2}\right)} , \quad
        \alpha _{\phi \phi } =  \frac{3r^{2}\left( 3a^{2} -r\mathcal{Z}\right)}{\mathcal{Z}^{2}},
        \end{aligned}
        \label{JMRFkds}
\end{equation}
and the corresponding one-form
\begin{equation}
    \beta _{\phi } = \frac{a(\mathcal{Z} +3r)}{\mathcal{Z}} .
    \label{betakds}
\end{equation}
Here $\mathcal{Z} =\Lambda r^{3} +\left( \Lambda a^{2} -3\right) r+6M$. In the next section, we will calculate the deflection angle for equatorial light in KdS spacetime within $M^{(\alpha 2)}_{kds}$.

\section{the corrected deflection angle} \label{CalDefAngle}
In this section, we calculate the deflection angle based on the revised orbit solution Eq.~\eqref{uofphi01}, the definition of the finite-distance deflection angle Eq.~\eqref{defineangle}, and the two-dimensional space $M^{(\alpha 2)}_{kds}$ Eq.~\eqref{malpha2kds}. The approach we employ is the generalized GW method for stationary spacetime proposed by Huang and Cao \cite{huang2023generalized}, which is more powerful than the original GW method \cite{gibbons2008applications}.

\begin{figure}[!ht]
    \centering
    \includegraphics[width=0.9\columnwidth]{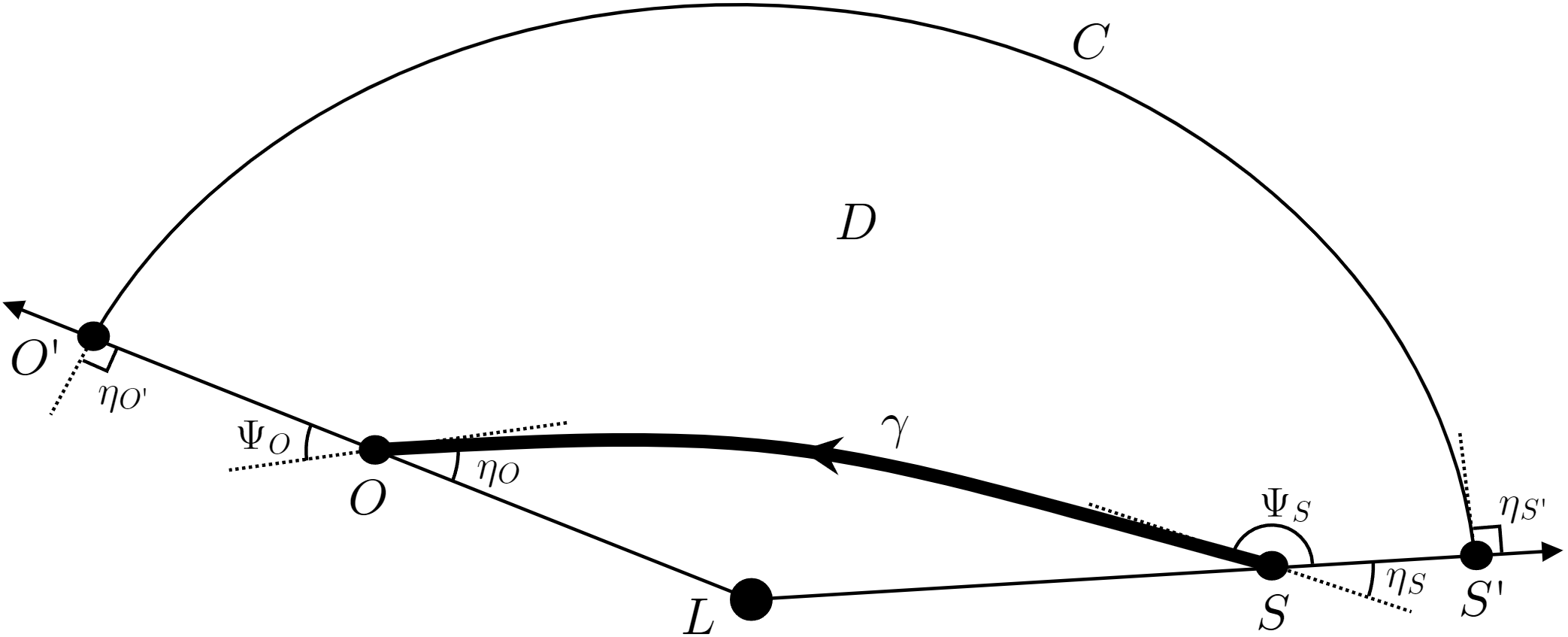}
    \caption{The quadrilateral region $D=^{O'}_{O}\square^{S'}_S$ in $M^{(\alpha 2)}_{kds}$.}
    \label{fig-3}
  \end{figure}

As shown in Fig.~\ref{fig-3},
in $M^{(\alpha 2)}_{kds}$, $\gamma$, $S$, $O$, $L$, $\Psi_S$ and $\Psi_O$ carry the same meaning as those illustrated in Fig.~\ref{fig-2}. $C=\overset{\curvearrowright}{S' O'}$ is an auxiliary circular arc with $r=r_c\in (r_\gamma^{max}, r_\Lambda)$, where $r_\gamma^{max}$ represents the maximal radial coordinate of $\gamma$ and $r_\Lambda$ denotes the cosmological horizon of KdS spacetime. $C$ intersects with the outward radial curves $\overrightarrow{LS}$ and $\overrightarrow{LO}$ at $S'$ and $O'$, respectively. $\eta_S$, $\eta_{S'}$, $\eta_{O'}$ and $\eta_O$ indicate the jump angles at $S$, $S'$, $O'$ and $O$, in that order. Then we get a quadrilateral region $D=^{O'}_{O}\square^{S'}_S$. 

Applying the Gauss-Bonnet theorem (pp. 272 and 277 in Ref.~\cite{manfredo1976carmo}) to $D$ brings about
\begin{equation}
    \begin{aligned}
        &\iint _{D} K\mathrm{d} S+\int _{\overrightarrow{SS} '} \kappa \mathrm{d} l+\int _{C} \kappa \mathrm{d} l+\int _{\overrightarrow{O' O}} \kappa \mathrm{d} l\\
        &+\int _{\overset{\curvearrowright }{OS}} \kappa \mathrm{d} l +\eta _{S} +\eta _{S'} +\eta _{O'} +\eta _{O} =2\pi \chi (D ).
    \end{aligned}
    \label{gbD}
  \end{equation}
where $K$, $\mathrm{d}S$ and $\chi\left(D\right)$ stand for the Gaussian curvature, area element and Euler characteristic number of $D$, respectively; $\kappa$ and $\mathrm{d}l$ denote the geodesic curvature and line element of curves, respectively. Substituting $\kappa(\overrightarrow{SS'})=\kappa(\overrightarrow{O'O})=0$ (refer to Appendix A of Ref.~\cite{huang2023finite} for the proof), $\int_{\overset{\curvearrowright}{OS}} \kappa \mathrm{d}l=-\int_\gamma \kappa \mathrm{d}l$, $\eta_S = \pi - \Psi_S$, $\eta_{S'}=\eta_{O'}=\pi/2$, $\eta_O=\Psi_O$, and $\chi\left(D\right)=1$ ($D$ is simply connected) into Eq.~\eqref{gbD} and combining the result with the definition Eq.~\eqref{defineangle}, the finite-distance deflection angle can be expressed as
\begin{equation}
    \delta = - \iint_{D} K \mathrm{d}S -\int_{C} \kappa \mathrm{d}l   +\phi_{OS}+ \int_\gamma \kappa \mathrm{d}l.
    \label{deltageneral}
\end{equation}
According to the derivation in Secs.~4.1 and 4.4 of Ref.~\cite{huang2023generalized}, the above formula becomes
\begin{equation}
    \delta = \int_{\phi_S}^{\phi_O} f(r_\gamma) \mathrm{d}\phi,
    \label{delta0}
\end{equation}
where $r_\gamma$ represents the radial coordinate of $\gamma$ and is expressed in terms of the azimuthal coordinate $\phi$ in calculation, $\phi_S$ and $\phi_O$ are the azimuthal coordinate of the source and observer, respectively, $f$ is defined by
\begin{equation}
    f(r)=1-\frac{\alpha_{\phi\phi,r}}{2\sqrt{\alpha_{rr}\alpha_{\phi\phi}}} -\beta_{\phi,r} \sqrt{ \frac{r^4}{\alpha_{\phi\phi}} \left( \frac{\mathrm{d}u}{\mathrm{d}\phi}\right)^2+ \frac{1}{\alpha_{rr}} }. 
    \label{fr}
\end{equation}

Substituting the metric of $M^{(\alpha 2)}_{kds}$ Eq.~\eqref{JMRFkds}, the corresponding one-form Eq.~\eqref{betakds} and the corresponding $\left(\mathrm{d}u/\mathrm{d}\phi\right)^2$ Eq.~\eqref{dudphikds} into Eq.~\eqref{fr}, then combing the result with the revised orbit solution of equatorial light in KdS spacetime, $r_\gamma$ (the reciprocal of Eq.~\eqref{uofphi01}), we have
\begin{equation}
    \begin{aligned}
        f(r_{\gamma } )= & M\cdot \frac{2\sin \phi }{b} -\Lambda \cdot \frac{b^{2}\csc^{2} \phi }{6} +M^{2} \cdot \frac{1}{4b^{2}}\Bigl[ 15\\
         & +\cos (2\phi )\Bigr] -Ma\cdot \frac{2\sin \phi }{b^{2}} +M\Lambda \cdot \frac{b\sin \phi }{3}\\
         & \cdot \Bigl( 1+2\csc^{4} \phi -\csc^{2} \phi \Bigr) +a\Lambda \cdot \frac{2b\csc^{2} \phi }{3}\\
         & -\Lambda ^{2} \cdot \frac{b^{4}\csc^{4} \phi }{72}\Bigl[ 2\cos (2\phi )+1\Bigr] +\mathcal{O}\left( \epsilon ^{3}\right) .
        \end{aligned}
\end{equation}
Denoting the indefinite integral of $f(r_\gamma)$ as $F(\phi)$, Eq.~\eqref{delta0} can be recast as $\delta = F\left(\phi_O\right)-F\left(\phi_S\right)$, namely
\begin{equation}
    \delta=F\left[\pi-\Phi\left(u_O\right)\right]-F\left[\Phi\left(u_S\right)\right],
\end{equation}
where Eq.~\eqref{phigamma} is used and $\phi_S<\pi/2<\phi_O$ is assumed. Finally, with the expression of $\Phi(u)$ Eq.~\eqref{PHIofu}, we obtain the finite-distance deflection angle of equatorial light in KdS spacetime in terms of the radial coordinate of the source and observer
\begin{equation}
    \begin{aligned}
        \delta = & M\cdot \frac{2}{b}\left(\sqrt{1-b^{2} u_{O}^{2}} +\sqrt{1-b^{2} u_{S}^{2}}\right)\\
         & -\Lambda \cdot \frac{b}{6}\left(\frac{\sqrt{1-b^{2} u_{O}^{2}}}{u_{O}} +\frac{\sqrt{1-b^{2} u_{S}^{2}}}{u_{S}}\right)\\
         & +M^{2} \cdot \frac{1}{4b^{2}}\Biggl\{15\bigl[\arccos (bu_{O} )+\arccos (bu_{S} )\bigr]\\
         & +\frac{15bu_{O} -7b^{3} u_{O}^{3}}{\sqrt{1-b^{2} u_{O}^{2}}} +\frac{15bu_{S} -7b^{3} u_{S}^{3}}{\sqrt{1-b^{2} u_{S}^{2}}}\Biggr\}\\
         & -Ma\cdot \frac{2}{b^{2}}\left(\sqrt{1-b^{2} u_{O}^{2}} +\sqrt{1-b^{2} u_{S}^{2}}\right)\\
         & +M\Lambda \cdot \frac{b}{6}\left(\frac{1}{\sqrt{1-b^{2} u_{O}^{2}}} +\frac{1}{\sqrt{1-b^{2} u_{S}^{2}}}\right)\\
         & +a\Lambda \cdot \frac{2}{3}\left(\frac{\sqrt{1-b^{2} u_{O}^{2}}}{u_{O}} +\frac{\sqrt{1-b^{2} u_{S}^{2}}}{u_{S}}\right)\\
         & -\Lambda ^{2} \cdot \frac{b}{72}\left(\frac{1-b^{2} u_{O}^{2} -2b^{4} u_{O}^{4}}{u_{O}^{3}\sqrt{1-b^{2} u_{O}^{2}}} +\frac{1-b^{2} u_{S}^{2} -2b^{4} u_{S}^{4}}{u_{S}^{3}\sqrt{1-b^{2} u_{S}^{2}}}\right)\\
         & +\mathcal{O}\left( \epsilon ^{3}\right).
        \end{aligned} \label{deltarst}
\end{equation}

We provide a brief discussion of Eq.~\eqref{deltarst}. For terms do not involving $\Lambda$, i.e., those corresponding to $M$, $M^2$ and $Ma$, they can be simplified under the infinite-distance limit of the source and observer, although it is unphysical for KdS spacetime. Specifically, when $u_S=u_O=0$, the first, third, and fourth term of Eq.~\eqref{deltarst} becomes $4M/b$, $15\pi M^2/(4b^2)$ and $-4Ma/b^2$, respectively. These three results are respectively consistent with the first, third, and fourth term of Eq.~\eqref{DAKdSSultana}, which is obtained by Sultana assuming the source and observer are positioned very far from the lens. While for terms involving $\Lambda$, namely those corresponding to $\Lambda$, $M\Lambda$, $a\Lambda$, and $\Lambda^2$, they will diverge under the infinite-distance limit, except for the $M\Lambda$ term.

\section{Conclusion}
\label{conclusion}
The introduction of the RI method aims to explore the influence of the cosmological constant on the light deflection. Using this approach, Sultana studied the deflection angle of equatorial light in KdS spacetime. However, in light of researchers' criticisms of the RI method, Sultana's finding inherently bears the limitations associated with this methodology.

In the context of KdS spacetime, we systematically address and rectify the orbit problem, position problem, and static problem identified in Sultana’s work. This is achieved by directly solving the original EOM, utilizing a finite-distance deflection angle, and adopting the Randers optical space, respectively. Building upon these rectification, we derive the finite-distance deflection angle of equatorial light in KdS spacetime with the generalized GW method. Our result is accurate up to second order with respect to the mass of the black hole $M$, the spin parameter of the black hole $a$ and the cosmological constant $\Lambda$.

As observing technologies advance for various celestial systems, the subtle differences between values given by different theoretical results are expected to be detected. The result presented in this paper, considering more detailed aspects and greater realism, undoubtedly provide a more accurate foundation for investigating the influence of the cosmological constant on light deflection.

\section*{Acknowledgments}
This work was supported in part by the National Natural Science Foundation of China Grant No. 12375045 and in part by the Science Research Fund of Hunan Provincial Education Department No. 21A0297.

\appendix

\bibliography{refs-kds}

\end{document}